\documentclass[aip,amsmath,amssymb,reprint]{revtex4-1}

\usepackage{graphicx}
\usepackage{dcolumn}
\usepackage{amssymb}
\usepackage{amsthm}
\usepackage{amsmath}
\usepackage{amsfonts}
\usepackage{algorithmic}
\usepackage{enumerate}
\usepackage{latexsym}
\usepackage{bm}
\usepackage{subfigure}
\usepackage[dvips]{color}
\usepackage[utf8]{inputenc}
\usepackage[T1]{fontenc}
\usepackage{mathptmx}
\usepackage{relsize}

\begin{document}

\preprint{AIP/123-QED}

\title[Chebyshev Polynomial Method]{Chebyshev Polynomial Method to Landauer-B\"{u}ttiker Formula of Quantum Transport in Nanostructures}

\author{Yan Yu}
\affiliation{SKLSM, Institute of Semiconductors, Chinese Academy of Sciences, P.O. Box 912, Beijing 100083, China}
\affiliation{School of Physical Sciences, University of Chinese Academy of Sciences, Beijing 100049, China}

\author{Yan-Yang Zhang}
\email{yanyang@gzhu.edu.cn}
\affiliation{School of Physics and Electronic Engineering, Guangzhou University, 510006
Guangzhou, China}

\affiliation{Research Center for Advanced Information Materials, Guangzhou University, 510006
Guangzhou, China}

\author{Lei Liu}
\affiliation{Department of Medical Physics, School of Medical Imaging, Hebei Medical University, Shijiazhuang, Hebei 050017, China}

\author{Si-Si Wang}
\affiliation{SKLSM, Institute of Semiconductors, Chinese Academy of Sciences, P.O. Box 912, Beijing 100083, China}
\affiliation{College of Materials Science and Opto-Electronic Technology, University of Chinese Academy of Sciences, Beijing 100049, China}

\author{Ji-Huan Guan}
\affiliation{SKLSM, Institute of Semiconductors, Chinese Academy of Sciences, P.O. Box 912, Beijing 100083, China}
\affiliation{School of Physical Sciences, University of Chinese Academy of Sciences, Beijing 100049, China}

\author{Yang Xia}
\affiliation{Microelectronic Instrument and Equipment Research Center, Institute of Microelectronics of Chinese Academy of Sciences, Beijing 100029, China}
\affiliation{School of Microelectronics, University of Chinese Academy of Sciences, Beijing 100049,
China}
\author{Shu-Shen Li}
\affiliation{SKLSM, Institute of Semiconductors, Chinese Academy of Sciences, P.O. Box 912, Beijing 100083, China}
\affiliation{College of Materials Science and Opto-Electronic Technology, University of Chinese Academy of Sciences, Beijing 100049, China}
\affiliation{Synergetic Innovation Center of Quantum Information and Quantum Physics, University of Science and Technology of China, Hefei, Anhui 230026, China}

\date{\today}

\begin{abstract}
Landauer-B\"{u}ttiker formula describes the electronic quantum transports in nanostructures and molecules. It will be numerically demanding for simulations of complex or large size systems due to, for example, matrix inversion calculations. Recently, Chebyshev polynomial method has attracted intense interests in numerical simulations of quantum systems due to the high efficiency in parallelization, because the only matrix operation it involves is just the product of sparse matrices and vectors. Many progresses have been made on the Chebyshev polynomial representations of physical quantities for isolated or bulk quantum structures. Here we present the Chebyshev polynomial method to the typical electronic scattering problem, the Landauer-B\"{u}ttiker formula for the conductance of quantum transports in nanostructures. We first describe the full algorithm based on the standard bath KPM. Then, we present two simple but efficient improvements. One of them has a time consumption remarkably less than the direct matrix calculation without KPM. Some typical examples are also presented to illustrate the numerical effectiveness.
\end{abstract}

\maketitle

\section{Introduction}
Landauer-B\"{u}ttiker formula plays an important role in the study of electronic quantum transports in nanostructures\cite{Landauer1957,Buettiker,TransmissionRMP,Datta}, molecular systems\cite{Modelular1}, and even DNAs\cite{DNA1}. It also plays an important role in calculating the thermal\cite{Thermal2,Thermal3,Thermal4}, optical\cite{Optical,Optical2} and phonon\cite{PhononTransport} transports in quantum structures. Landauer-B\"{u}ttiker formula relates the electronic conductance of a two-terminal or multi-terminal device to the quantum transmission\cite{Landauer1957,Buettiker,TransmissionRMP}. The quantum transmission can be expressed in terms of Green's functions, which is a standard numerical tool today\cite{Datta,QuantumTransport,Kwant,MoS2Ribbons}. Since this is a real space method,
it is computationally demanding for a system related with large number of orbital basis, e.g., large size systems\cite{LargeGraphene,LimitQuantumTransport}, biological molecules\cite{DNA1} and (quasi-)incommensurate systems\cite{Amorphous,QuasiCrystal,commensurate1}.

In the recent decade, a powerful numerical method treating with Hamiltonians on large Hilbert spaces has attracted attention, the kernel polynomial method (KPM), such as the Chebyshev expansion\cite{KPMRMP}. In most KPM calculations, the only matrix operations involved are product between sparse matrices (Hamiltonians) and vectors, and matrix traces. For a sparse matrix with dimension $D$, the matrix vector multiplication is only an order $O(D)$ process. Thus the calculation of $N$ moments of Chebyshev terms needs $O(ND)$ operations and time\cite{KPMRMP}. A direct application of KPM is the calculation of the spectral function of an isolated system\cite{KPMRMP,SpectralFunctionKMP1,SpectralFunctionKMP2}. Taking advantage of appropriate analytical continuation, one can arrive at the evaluation of Green's functions\cite{KPMRMP,GreenFunctionKMP}. Expressions of physical quantities in terms of KPM have been developed recently\cite{LocLengthKMP,ResponseFunctionKMP,DynCorr,SJYuan2010,SJYuan2010PRL,SJYuan2011,SJYuan2016,KuboFormula,ZYFan2018}, including the applications to superconductors\cite{BdGKPM}, topological materials\cite{RealSpace,LeiLiu2018}, quantum impurity problems\cite{Impurity1,Impurity2,SelfEnergyKMP} and \emph{ab initio} calculations\cite{Abinitio0,Abinitio}. However, these methods are applicable to bulk or isolated systems\cite{KPMRMP,ZYFan2018}, not to scattering processes between leads in open systems, which corresponds to a realistic experimental setup\cite{Datta}.

In this paper, we will propose some KPM methods to calculate the
Landauer-B\"{u}ttiker transmission in a two-terminal system: a conductor connected to left (L) and right (R) leads, as illustrated in Fig. \ref{FigDevice}. The transmission through the conductor can be written in terms of Green's functions, where the leads manifest themselves as self energies\cite{Datta}. This is typical context of an open system coupled to a bath\cite{BathKMP,FermionicBath}. We first fully describe this problem as a generalization of the standard bath technique of
KPM\cite{BathKMP}, where the needed self energies and dressed Green's function as Chebyshev polynomials of some sparse matrices. To reduce the numerical consumption, we then propose two practical improvements. One of them can largely simplify the self-consistent calculation of the self energies, and the second of them can even avoid this self-consistent process, which has a much less time and space consumption than those of the direct method of matrix evaluation without KPM.

This paper is organized as follows. After the general introduction, in Sections II and III, we briefly introduce the basic knowledge of Chebyshev polynomials and Landauer-B\"{u}ttiker formula, respectively. In Section IV, we describe the algorithm of calculating Landauer-B\"{u}ttiker formula with Chebyshev polynomials, including the calculation of dressed Green's function and lead self energies, following the standard bath method of KPM. To reduce the numerical demanding, we propose two practical improvements in Section V. Some numerical examples are presented in Section VI. In Section VII, we provide a summary and some outlooks for future works.

\section{Chebyshev Expansion and the Kernel Polynomial Method}

In this section, we briefly summarize the definition and  basic properties of Chebyshev polynomials that will be used. The Chebyshev polynomials $T_{n}(x)$ with $x\in[-1,1]$ are in the explicit form as\cite{KPMRMP}
\begin{equation}\label{equ:A1}
T_{n}(x)=\cos[n\arccos(x)]\;,
\end{equation}
which satisfy the recursion relations,
\begin{equation}\label{EqRecursion}
\begin{split}
& T_{0}(x)=1\;,T_{-1}(x)=T_{1}(x)=x\;\\
& T_{n+1}(x)=2xT_{n}(x)-T_{n-1}(x)\;.
\end{split}
\end{equation}
The scalar product is defined as
\begin{equation}\label{EqScalarProduct}
\langle T_{m}|T_{n}\rangle\equiv \int_{-1}^1 \frac{T_{m}(x)T_{n}(x)}{\pi\sqrt{1-x^{2}}},
\end{equation}
with the weight function $(\pi\sqrt{1-x^{2}})^{-1}$. It is thus easy to verify the orthogonality relation between
Chebyshev polynomials,
\begin{equation}\label{EqOrthogonality}
\langle T_{m}|T_{n}\rangle=\frac{1+\delta_{n,0}}{2}\delta_{n,m}.
\end{equation}

In terms of these orthogonality relations (\ref{EqOrthogonality}), a piecewise smooth and continuous function $f(x)$ with $x\in [-1,1]$ can be expanded as
\begin{equation}\label{equ:A5}
f(x)=\frac{1}{\pi\sqrt{1-x^{2}}}\left[\mu_{0}+2\sum_{n=1}^{\infty}\mu_{n}T_{n}(x)\right],
\end{equation}
with expansion coefficients $ \mu_{n}$
\begin{equation}\label{equ:A6}
\mu_{n}=\int^{1}_{-1}f(x)T_{n}(x)dx
\end{equation}

Practically, the function $f(x)$ should be numerically reconstructed from a truncated series with the first $N$ terms in Eq. (\ref{equ:A5}). However, experiences show that the numerical performance of this simple truncation is bad, with slow convergence and remarkable fluctuations (Gibbs oscillations)\cite{KPMRMP}. This can be improved by a modification of the expansion coefficients as $\mu_n\rightarrow g_n \mu_n$, where $\{ g_n \}$ is the kernel. In other words, appropriate choices of the kernel $\{ g_n \}$ will make the truncated series a numerically better approximation of the function\cite{KPMRMP}
\begin{equation}\label{equ:A7}
f(x)\!\approx\! f_{\mathrm{KPM}}(x)\!=\!\frac{1}{\pi\sqrt{1-x^{2}}}\!\left[g_{0}\mu_{0}\!+\!2\sum_{n=1}^{N-1}g_{n}\mu_{n}T_{n}(x)\right],
\end{equation}
Among different kernels, here we adopt the Jackson kernel with the explicit expression as\cite{KPMRMP}
\begin{equation}\label{equ:A8}
\mathlarger{g}_{n}^{J}=\frac{(N-n+1)\cos\frac{\pi n}{N+1}+\sin\frac{\pi n}{N+1}\cot\frac{\pi}{N+1}}{N+1},
\end{equation}
which is suitable for the applications related to Green's functions.

Besides a numeric function $f(x)$, the Chebyshev expansion can also be used to approximate the function of a Hermitian operator $H$ (or equivalently its matrix $\bm{H}$ in an appropriate representation), if the eigenvalue spectrum of $H$ is within the interval $[-1,1]$\cite{KPMRMP}. For a general Hermitian operator, e.g. a Hamiltonian $H$ with maximum (minimum) eigenvalue $E_{\mathrm{max}}$  ($E_{\mathrm{min}}$), this condition of spectrum can be satisfied by simply performing an appropriate rescaling on the matrix (and also on the energy scale),
\begin{equation}\label{equ:A9}
\tilde{H}=\frac{1}{a}\big(H-b\big),\qquad\tilde{E}=\frac{E-b}{a}
\end{equation}
with
\begin{equation}\label{equ:A10}
a=\frac{E_{\max}-E_{\min}}{2-\zeta},\qquad b=\frac{E_{\max}+E_{\min}}{2},
\end{equation}
so that the the spectrum of $\tilde{H}$ is within $[-1,1]$. Here the parameter $\zeta>0$ is a small cutoff to avoid numerical instabilities at the boundaries $\pm 1$. A proper rescaling, i.e., an appropriately small $\zeta$ will reduce the necessary $N$ for a certain expansion to reach the same precision. Throughout this work, we fix $\zeta=0.01$. In practical uses, the lower and upper bounds of $\bm{H}$ can be estimated by using sparse matrix eigenvalue solvers, e.g., the FEAST algorithm of Intel MKL. After the calculation of physical properties with the help of Chebyshev polynomials,
their correct dependence on the energy $E$ can be restored by a simple inverse transformation of Eq. (\ref{equ:A9}). Therefore in the following, we will always consider that the operator matrices have been rescaled according to Eq. (\ref{equ:A9}) before they enter Chebyshev polynomials, and the tilde hats on the operators and eigenvalues will be omitted.
It can be shown that\cite{KPMRMP,RealSpace,GreenFunctionKMP}, the retarded (advanced) Green's function
\begin{equation}\label{EqBareGreenFunction}
G^{r(a)}(E,H)=\lim_{\eta \to 0+}\big[(E\pm i\eta)I-H\big]^{-1}
\end{equation}
at energy $E$ can be expanded in terms of Chebyshev kernel polynomials as
\begin{equation}\label{EqChebyshevGreen}
G^{r(a)}(E,H)=\frac{i}{\sqrt{1-E^{2}}}(g_{0}\mu_{0}+2\sum_{n=1}^{N-1}g_{n}\mu_{n}e^{\mp in\arccos(E)})\,
\end{equation}
with coefficient matrices
\begin{equation}\label{EqChebyshevCoeffH}
\mu_{n}=\mp T_{n}(H).
\end{equation}
Now the broadening $\eta$ does not explicitly appear in the matrix elements. Rather, it is associated with $N$, the number of expansion moments. Larger $N$ corresponds to a smaller $\eta$. Notice $G$ and $\mu_{n}$ are also operators. In a certain representation, these operators can be explicitly written as corresponding matrices $\bm{H}$, $\bm{G}$ and $\bm{\mu}_n$ with the same size.
Throughout this manuscript, all matrices will be written in a bold form of the corresponding operator.

\section{Electronic Transmission in Terms of Green's Functions}

\begin{figure}[htbp]
\centering
\includegraphics*[width=0.5\textwidth]{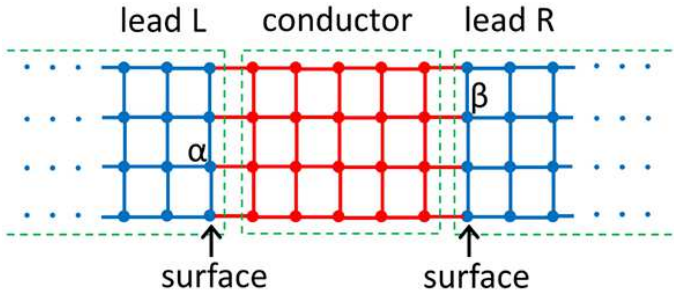}
\centering
\caption{The typical setup of a two-terminal measurement of quantum transport, a conductor (red) is connected to left (L) and right (R) leads (blue). Sites $\alpha$ and $\beta$ are on the surfaces of the leads, which are adjacent to the conductor. }
\label{FigDevice}
\end{figure}

In this section, we briefly review the Landauer-B\"{u}ttiker formula represented as Green's functions. Consider the two-terminal transport device illustrated in Fig. \ref{FigDevice},
with one conductor connected to two semi-infinite leads.
Formally, the Hamiltonian of this combined system can be written as
\begin{equation}\label{EqFullHamiltonian}
H=H_{C}+H_{L}+H_{R}+H_{CL}+\mathrm{H. c.}+H_{CR}+\mathrm{H. c.},
\end{equation}
where $H_{C}$ is the Hamiltonian of the conductor, $H_{L}$ ($H_{R}$) is that of the left (right) lead,
and $H_{CL}$ ($H_{CR}$) is the coupling from the conductor to the left (right) lead.
It is convenient to write these Hamiltonians in the real space representation (tight binding model) as matrices.  For example, the real space Hamiltonian of a conductor (lead) can be expressed in a generic second quantization form as
\begin{equation}\label{EqHamiltonianMatrix}
H=\sum_{\alpha,\beta} \bm{H}_{\alpha\beta}c_{\alpha}^{\dagger}c_{\beta},
\end{equation}
with $c_{\beta}$ the annihilation operator of the spinorbital $\beta$ in the conductor (lead). Here $\bm{H}$ is a matrix with elements $\bm{H}_{\alpha\beta}$.

Due to the coupling to leads, now the (retarded) Green's function of the conductor $G^{r}_{C}$
is, of course, not the original bare one $\big[E + i\eta - H_{C}\big]^{-1}$. Thanks to the Dyson equation of Green's functions, it can be expressed as the dressed one as\cite{Datta,DHLee}
\begin{equation}\label{EqDressedGreenFunction2}
G^{r}_{C}(E)=\big[E - H_C-\Sigma_{L}(E)-\Sigma_{R}(E)\big]^{-1},
\end{equation}
where $\Sigma_{L}$ ($\Sigma_{R}$) is the self energy of the left (right) lead.
The technique of self energy liberate one from inserting the full Hamiltonian (\ref{EqFullHamiltonian})
into Eq. (\ref{EqBareGreenFunction}) to obtain the dressed Green's function of the conductor.
The self energy is the result of integrating out the degree of freedom of the lead\cite{Datta,CMFT}, i.e.,
\begin{equation}\label{EqSelfEnergy}
\Sigma_{p}(E)=H^{\dagger}_{pC}G^{r}_{p}(E)H_{pC},\quad p\in\{L,R\}
\end{equation}
where $G^{r}_{p}$ is the Green's function of lead $p$, and $H_{pC}$ is the coupling Hamiltonian between lead $p$ and the conductor.
In the real space representation, $\bm{G}^{r}_{p}$  is an infinite dimensional matrix because the lead is semi-infinite.
However, since only a few spinoribitals of the lead is connected to the conductor through $H_{pC}$, in the evaluation of Eq. (\ref{EqSelfEnergy}), we only need to know the ``surface'' subset of the matrix $\bm{G}^{r}_{p}$, i.e., those matrix elements $\big[\bm{G}^{r}_{p}\big]_{\alpha\beta}$ with $\alpha$ and $\beta$ running over spinorbitals connected to the conductor.
This subset will be called the surface Green's function.

At zero temperature, the two-terminal conductance $G$ in Fig.\ref{FigDevice} is
represented as the Landauer-B\"{u}ttiker formula\cite{Landauer1957,Buettiker,TransmissionRMP,Datta},
\begin{equation}
G=\frac{e^2}{h}T,
\end{equation}
where $e$ is the elementary charge, $h$ is the Planck constant, and $T$
is the transmission through the conductor.
This transmission $T$ at Fermi energy $E$ can be expressed in terms of Green's functions as\cite{Datta,QuantumTransport},
\begin{equation}\label{EqnTransmission}
T(E)=\mathrm{Tr}[\Gamma_{R}(E)G_C^{r}(E)\Gamma_{L}(E)G_C^{a}(E)],
\end{equation}
where
\begin{equation}\label{EqDressedGreenFunction}
G^{a}_C(E)=\big[G^{r}_C(E)\big]^{\dagger},\qquad \Gamma_{L(R)}=i\big[\Sigma_{L(R)}-\Sigma_{L(R)}^{\dagger}\big]
\end{equation}

Traditionally, the self energies (\ref{EqSelfEnergy}) of the leads can be calculated explicitly by a direct diagonalization method\cite{DHLee} or an iterative method\cite{InterativeSurfaceGreen}. Afterwards, they are inserted into Eqs. (\ref{EqDressedGreenFunction2}), (\ref{EqDressedGreenFunction}) and finally (\ref{EqnTransmission}) for the evaluation of the transmission. In this process, the most time-consuming step will be the calculation of lead self energies, and the matrix inversion (which does not preserve the sparseness of the matrix) in Eq. (\ref{EqDressedGreenFunction2}). For a two-terminal device simulation where the conductor lattice can be well divided into layers of sites (layers should be defined in such a way that hoppings only exist between nearest layers), the simulation can be decomposed into a layer-to-layer recursive method, which is based on the Dyson equation for Green's functions\cite{QuantumTransport,Recursive}. This decomposition can remarkably reduce the time and space consumption in calculations. However, this recursive method will be technically tedious for a multi-terminal setup, and even impossible for, say, a twisted bilayer graphene\cite{TwistedGraphene1,TwistedGraphene2}. In these examples, one still needs to calculate the full-size and dense matrices associated with Hamiltonians and Green's functions directly. In the following, we will investigate algorithms based on KPM to calculate Eq. (\ref{EqnTransmission}), with slightly different steps.

\section{Standard Bath Chebyshev Polynomial Method}

Before evaluating the transmission function (\ref{EqnTransmission}) from the Hamiltonian (\ref{EqFullHamiltonian}), two steps are essential: First, solving the self energies [Eq. (\ref{EqSelfEnergy})]; Second, inclusion of them into
the conductor's Green's function as Eq. (\ref{EqDressedGreenFunction2}).
The numerical treatments of these steps by direct matrix calculations have been very mature and well-known\cite{Datta,QuantumTransport}.
However, in the context of KPM, the realization of these steps is not easy nor straightforward,
especially if one insists to avoid calculations related to large dense matrices.
We achieve this goal by a generalization of the bath technique of KPM\cite{BathKMP},
which will be described here in detail. In Section V B, another distinct algorithm will be introduced.

The lead connected to the conductor is semi-infinitely long
and therefore it can be viewed as a bath\cite{BathKMP}. The central task of obtaining the lead self energy [Eq. (\ref{EqSelfEnergy})] is to calculate the surface Green's function $\big[\bm{G}^{r}_{p}\big]_{\alpha\beta}$ of lead $p$, with $\alpha$ and $\beta$ running over the surface which will be connected to the conductor. In the context of KPM method, we need to calculate the Chebyshev coefficient matrix $\bm{\mu}_n^{\alpha\beta}$ in Eq. (\ref{EqChebyshevGreen}) of the lead. This, of course, cannot be calculated by using Eq. (\ref{EqChebyshevCoeffH}) directly, as the lead Hamiltonian matrix $\bm{H}_p$ is infinite dimensional. Instead, we will use a self consistent method as described below.

\subsection{Basic Definitions}
First, some useful mathematical structures related to an isolated lead $p$ will be constructed. As suggested in Ref. \cite{BathKMP}, we define the Chebyshev vectors as
\begin{equation}\label{EqChebyshevVector}
|n_{\alpha}\rangle \equiv T_{n}(H_{p})f_{\alpha}^{\dagger}|\mathrm{vac}\rangle, (n\in\mathbb{N})
\end{equation}
with $|\mathrm{vac}\rangle$ describing the lead vacuum, i.e.,  $f_{\alpha}^{\dagger}|\mathrm{vac}\rangle=0$, and $f_{\alpha}^{\dagger}$ the creation operator in the lead at spinorbital state $\alpha$. These Chebyshev vectors are not orthonormal and the scalar product
\begin{equation}\label{equ:A18}
\langle 0_{\beta}|n_{\alpha}\rangle =\langle \mathrm{vac}|f_{\beta}T_{n}(H_{p})f_{\alpha}^{\dagger}|\mathrm{vac}\rangle=\bm{\mu}_{n}^{\beta\alpha}.
\end{equation}
By comparing with Eq. (\ref{EqChebyshevCoeffH}), one can see that, this matrix $\bm{\mu}_{n}$ is just the $n$-th Chebyshev coefficient matrix of the lead's Green's function. The series of the Chebyshev vectors defined in Eq. (\ref{EqChebyshevVector}) span a Hilbert space $\mathcal{H}_{\alpha}$. As can be seen from the definition, $\mathcal{H}_{\alpha}$ is a subspace of the Fock space for the lead operator $f_{\alpha}^{(\dagger)}$. From the recursion relation, Eq. (\ref{EqRecursion}), it is easy to conclude the operation of $H_{p}$ on $\mathcal{H}_{\alpha}$ as
\begin{equation}
H_{p}|n_{\alpha}\rangle=
\begin{cases}
|1_{\alpha}\rangle,&n=0\\
(1/2)(|(n-1)_{\alpha}\rangle+|(n+1)_{\alpha}\rangle),&n>0.
\end{cases}
\end{equation}
In other words, in the subspace $\mathcal{H}_{\alpha}$, the effect of $H_{p}$ can be expressed in a matrix form as
\begin{equation}\label{equ:A20}
(\widehat{\bm{H}}^{\alpha}_{p})_{mn}\equiv \frac{1}{2}\left(
\begin{array}{ccccc}
0&1&0&0&\cdots\\
2&0&1&0& \\
 &1&0&1& \\
 & &1&0& \\
 & &\vdots& &\ddots
\end{array}
\right),
\end{equation}
with
\begin{equation}\label{EqHpmn}
H_{p}|n_{\alpha}\rangle=\sum_{m}(\widehat{\bm{H}}^{\alpha}_{p})_{mn}|m_{\alpha}\rangle.
\end{equation}
Notice that $(\widehat{\bm{H}}^{\alpha}_{p})_{mn}\neq\langle m_{\alpha}|H_{p}|n_{\alpha}\rangle$ owing to the non-orthogonality of these Chebyshev vectors.
For a truncation with $N$ Chebyshev terms (\ref{equ:A7}), the size of the matrix $\widehat{\bm{H}}^{\alpha}_{p}$ is $N \times N$.

Following Eq. (\ref{EqChebyshevVector}), another useful relation can be obtained as
\begin{eqnarray}\label{equ:A26}
f_{\beta}|n_{\alpha}\rangle&=f_{\beta}T_{n}(H_{p})f_{\alpha}^{\dagger}|\mathrm{vac}\rangle\\
&=|\mathrm{vac}\rangle\langle\mathrm{vac}|f_{\beta}T_{n}(H_{p})f_{\alpha}^{\dagger}|\mathrm{vac}\rangle=\bm{\mu}_{n}^{\beta\alpha}|\mathrm{vac}\rangle.
\end{eqnarray}

\subsection{Dressed Green's Function}

Suppose the Chebyshev coefficient matrices $\bm{\mu}_n$ of lead $p$ have been known. Now we connect a conductor $C$ to the lead. The Hamiltonian $H_C$ of the conductor is in the form of Eq. (\ref{EqHamiltonianMatrix}),
and the size of the corresponding matrix $\bm{H}_C$ is $M\times M$.
Without loss of generality, we consider the conductor-lead coupling
to be the following simple form
\begin{equation}\label{equ:A16}
H_{Cp} + \mathrm{H.c.} =\sum_{\alpha=1}^{W}t_{\alpha}(c_{\alpha}^{\dagger}f_{\alpha}+f_{\alpha}^{\dagger}c_{\alpha}),
\end{equation}
where $W$ is the effective ``width'' of the cross section, $c_{\alpha}$ ($f_{\alpha}$) is the annihilation operator in the conductor (lead), and $t_{\alpha}$ denotes the hopping matrix elements. As in most practical cases, we have considered these $W$ hopping bonds are coupling sites between the conductor and the lead in a one-to-one way.

Based on above definitions, now we can approximately express the full Hamiltonian $H_C+H_p+H_{Cp} + \mathrm{H.c.}$ in the finite-dimensional representation with basis ordered as
\begin{eqnarray}\label{EqCoupledMatrix}
&\Big( |1\rangle, \cdots,|\beta\rangle,\cdots|M\rangle, |0_{1}\rangle,\cdots,|n_{1}\rangle,\cdots,|N-1_{\, 1}\rangle,\nonumber\\
&\cdots,|n_{\alpha}\rangle,\cdots, |0_{W}\rangle,\cdots,|n_{W}\rangle,\cdots,|N-1_{\,W}\rangle \Big),
\end{eqnarray}
where $\huge|\beta\rangle$ ($1\leq\beta\leq M$) are spinorbital basis states in the conductor, and
$|n_{\alpha}\rangle$ ($0\leq n\leq N-1$ and $1\leq \alpha\leq W$) are Chebyshev vectors of lead $p$ defined in Eq. (\ref{EqChebyshevVector}). It can be easily shown that, the full Hamiltonian in this representation is a $(M+W\times N)$-dimensional sparse matrix with nonzero blocks illustrated as follows:
\begin{equation}\label{EqR}
\bm{R}=\left(
\begin{array}{ccc|cccccc}
\quad& & & & & & & &\\
\quad&  \text{\Large$ \bm{H}_C $}& & & & & & & \\
 \quad& & &\bm{L}^{1}_p&\bm{L}^{2}_p&\cdots&\bm{L}^{\alpha}_p&\cdots&\bm{L}^{W}_p\\ \hline
& & \bm{M}^{1}_p&\bm{\widehat{H}}^{1}_p& \\
& & \bm{M}^{2}_p&     &\bm{\widehat{H}}^{2}_p& \\
& & \vdots&    &     &\ddots& \\
& & \bm{M}^{\alpha}_p&&     &      &\bm{\widehat{H}}^{\alpha}_p& \\
& & \vdots&    &     &      &          &\ddots& \\
& & \bm{M}^{W}_p&     &     &      &          &      &\bm{\widehat{H}}^{W}_p\\
\end{array}
\right)
\end{equation}
Here, $\bm{H}_C$ is the Hamiltonian matrix of the isolated conductor with size $M \times M$, and $\bm{\widehat{H}}^{\alpha}_p$ are $N\times N$ matrices defined in Eq. (\ref{EqHpmn}) associated with lead $p$. As for the conductor-lead coupling sub-matrices, each $\bm{L}^{\alpha}_p$ is a $W\times N$ matrix in the form as
\begin{equation}\label{EqLalpha}
\bm{L}^{\alpha}_p=\left(
             \begin{array}{cccc}
               t_{\alpha}\mu_{0}^{1\alpha} & t_{\alpha}\mu_{1}^{1\alpha} & \cdots & t_{\alpha}\mu_{N-1}^{1\alpha} \\
               t_{\alpha}\mu_{0}^{2\alpha} & t_{\alpha}\mu_{1}^{2\alpha} & \cdots & t_{\alpha}\mu_{N-1}^{2\alpha} \\
                &  & \cdots &  \\
               t_{\alpha}\mu_{0}^{W\alpha} & t_{\alpha}\mu_{1}^{W\alpha} & \cdots & t_{\alpha}\mu_{N-1}^{W\alpha} \\
             \end{array}
           \right),
\end{equation}
where $t_{\alpha}$ are the coupling hopping integral defined in Eq (\ref{equ:A16}),
and $\bm{\mu_{n}}^{\beta\alpha}$ of the lead are defined in Eq. (\ref{equ:A18}), with $\{ \alpha,\beta \}$ only running over the surface spinorbital states connecting with the conductor. On the other hand, each $M^{\alpha}_p$ is an $N\times W$ matrix with only one nonzero element,

\begin{eqnarray}
\big(\bm{M}^{\alpha}_p \big)_{n\gamma}= \begin{cases}
t_{\alpha},&n=1 \quad\mathrm{and}\quad \gamma=\alpha \\
0,& \mathrm{otherwise}.
\end{cases}
\end{eqnarray}

This matrix $\bm{R}$ is determined if $\bm{H}_C$, $\bm{H}_{Cp}$ and $\bm{\mu}_{n}^{\beta\alpha}$ are known, and it plays the central role in the Chebyshev approach to quantum systems coupled to a bath\cite{BathKMP}.
It has been shown that\cite{BathKMP}, the dressed Green's function $\big[(E+i\eta)I-H_C-\Sigma_p\big]^{-1}$ of the conductor coupled to lead $p$ can be obtained by using Eq. (\ref{EqChebyshevGreen}) and Eq. (\ref{EqChebyshevCoeffH}), with $H$ replaced by $R$\cite{BathKMP}.

\subsection{Self Energy}

In most applications, the coefficients $\mu_{n}^{\beta\alpha}$ associated with the lead surface is not known a priori, and practically they cannot be obtained by using Eq. (\ref{equ:A18}). However, the above bath method also offers a practical algorithm of calculating the lead coefficients $\bm{\mu}_{n}^{\beta\alpha}$ in a self-consistent way, which will be described as follows.
\begin{figure}[htbp]
\centering
\includegraphics*[width=0.4\textwidth]{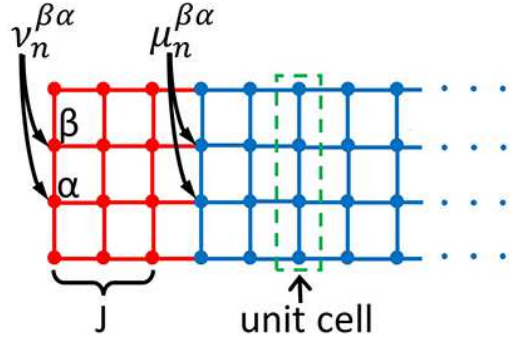}
\centering
\caption{The self-consistent calculation of the lead Chebyshev coefficients. A natural extension with $J$ unit cells (red) are connected the terminal of the original semi-infinite lead (blue). }
\label{FigSemiinfinitelead}
\end{figure}
The lead is a semi-infinite crystal whose one-dimensional unit cell can be defined in a natural way. In Fig. \ref{FigSemiinfinitelead}, we illustrate a lead extending infinitely to the right direction, with the unit cell marked by the green dashed frame. The lead Chebyshev coefficients of the left surface are $\bm{\mu}_{n}^{\beta\alpha}$, with $\alpha$ and $\beta$ only running over the left surface spinorbitals which will be connected to the conductor. Now we couple $J$ unit cells to the left surface of this lead. Then the Chebyshev coefficients $\bm{\nu}_{n}^{\beta\alpha}$ of the new left surface can be calculated by using the process introduced above. On the other hand, because these unit cells are just a natural extension of the lead, this new composite system ($J$ unit cells coupled to the original lead) is also semi-infinite and is essentially equivalent to the original lead. As a result, the self-consistent condition
\begin{equation}\label{EqSelfConsistent}
\bm{\nu}_{n}^{\beta\alpha}=\bm{\mu}_{n}^{\beta\alpha}
\end{equation}
should hold.

Practically, we can start from a guess of the lead Chebyshev coefficients, e.g., $\bm{\mu}_{n}^{\beta\alpha}=0$, and then repeatedly couple $J$ unit cells to the left surface of the lead and calculate the Chebyshev coefficients $\bm{\nu}_{n}^{\beta\alpha}$
associated with the new surface, until the self-consistent condition (\ref{EqSelfConsistent}) are satisfied within a given error. Although the choice $\bm{\mu}_{n}^{\beta\alpha}=0$ fail to meet the rule $\langle0_{\alpha}|0_{\alpha}\rangle=\langle\mathrm{vac}|f_{\alpha}f_{\alpha}^{\dagger}|\mathrm{vac}\rangle=\bm{\mu}_{0}^{\alpha\alpha}=1$, it does not affect the final convergence. A larger number $J$ of unit cells will consume more time for each iteration step, but will reduce the number of iteration steps. Therefore an appropriate $J$ should be carefully chosen for a concrete model. After the lead Chebyshev coefficients $\bm{\mu}_{n}^{\beta\alpha}$ are known, the surface Green's function can be obtained through Eq. (\ref{EqChebyshevGreen}), and the self energy through Eq. (\ref{EqSelfEnergy}).

\subsection{Counting Both Leads in}

So far, we have shown that, replacing $H$ in Eq. (\ref{EqChebyshevCoeffH}) with $\bm{R}$ defined in Eq. (\ref{EqR}) will give rise to the dressed Green's function of the conductor when coupled to a \emph{single} lead $p$, $\bm{G}_{C}=\big[E\bm{I}-\bm{H}_C-\bm{\Sigma}_p\big]^{-1}$.
For a two-terminal device, the inclusion of left (L) and right (R) leads
\begin{equation}\nonumber
\bm{G}_{C}=\big[E\bm{I}-\bm{H}_C-\bm{\Sigma}_L-\bm{\Sigma}_R\big]^{-1}
\end{equation}
can be similarly achieved by a trivial generalization of the matrix in Eq. (\ref{EqR}) as
\begin{equation}\label{EqR2}
\bm{R}=\left(
  \begin{array}{ccc|ccc|ccc}
    \bm{\widehat{H}}^{1}_L &  &  & \bm{M}_{L}^1 &  &  &  &  &  \\
      & \ddots &   & \vdots &   &   &   &   &   \\
      &   & \bm{\widehat{H}}^{W}_L & \bm{M}_{L}^{W} &   &   &   &   &   \\ \hline
    \bm{L}_{L}^{1} & \cdots & \bm{L}_{L}^W &   &   &   &   &   &   \\
      &   &   &  & \text{\Large$ \bm{H}_C $}   &   &   &   &   \\
      &   &   &   &   &   & \bm{L}_{R}^1 & \cdots & \bm{L}_{R}^W \\ \hline
      &   &   &   &   & \bm{M}_{R}^1 & \bm{\widehat{H}}^{1}_R &   &   \\
      &   &   &   &   & \vdots &   & \ddots &   \\
      &   &   &   &   & \bm{M}_{R}^W &   &   & \bm{\widehat{H}}^{W}_R \\
  \end{array}
\right),
\end{equation}
once the Chebyshev coefficients $\bm{\mu}_n$ associated with both leads were known.

\section{Practical Improvements}

\begin{figure}[htbp]
\centering
\includegraphics*[width=0.45\textwidth]{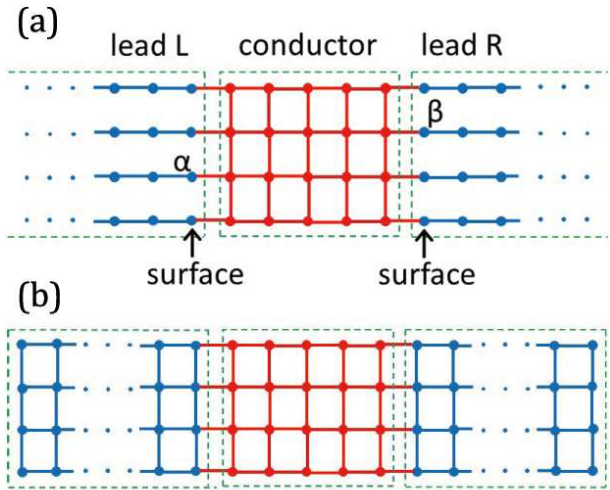}
\centering
\caption{ Two setups to simplify the calculations. (a) The leads (blue) are set to be decoupled 1D atomic chains. (b) The leads are set to have finite lengths $N_x^{L}$ and $N_x^{R}$, instead of being semi-infinitely long. }
\label{FigImprovements}
\end{figure}

\begin{figure*}[htbp]
\centering
\includegraphics*[width=1.0\textwidth]{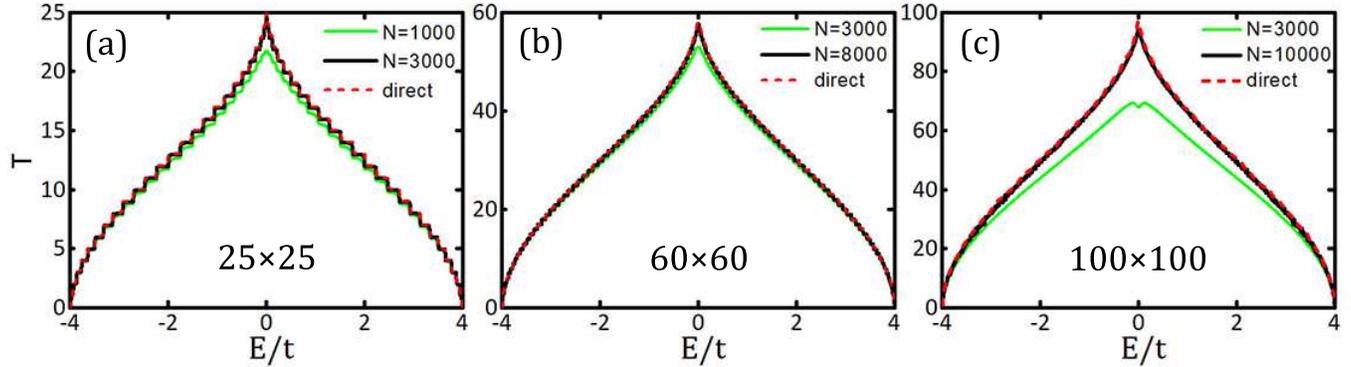}
\centering
\caption{ Results for Example A, square lattice conductor with square lattice leads by using the standard bath KPM. Here we present the transmission $T$ as a function of energy $E$, with leads the same width as the conductor, for different conductor sizes: (a) $25\times 25$, (b) $60\times60$ and (c) $100\times 100$. Solid lines are results from the standard bath KPM introduced in Section IV, with different Chebyshev terms $N$. Red dashed lines are the result of a direct matrix evaluation of Eq. (\ref{EqnTransmission}) without KMP, as a reference. Notice different scales of the longitudinal axes in different panels.}
\label{FigDifference}
\end{figure*}

The central task of KPM is to obtain the corresponding Chebysheve coefficient matrices. One merit of the KPM is that these Chebyshev coefficients are independent of the energy $E$, i.e., the transport properties over the full energy spectrum are known if the corresponding Chebyshev coefficients have been calculated out. Particularly, when plotting Fig. \ref{FigDifference} , one only needs to calculate the lead Chebyshev coefficients $\bm{\mu}_n^{\beta \alpha}$ for \emph{once}, and then the energy dependence enters simply through Eq. (\ref{EqChebyshevGreen}) which is numerically cheap. In the traditional matrix inversion method, on the other hand, the full process of calculating the self energy (\ref{EqSelfEnergy}) and the dressed Green's function (\ref{EqDressedGreenFunction}) should be performed separately for different energies, which are numerically independent.

The algorithm described in Section IV was based on a mathematically rigorous realization of the standard bath approach of the KPM\cite{BathKMP}, which is referred as the ``standard bath KPM''. In the practical simulations, however, calculating Chebysheve coefficient matrices from this standard method might be very numerically demanding on central
Q4 processing unit (CPU) based computers.
For example, in the calculation of conductance in terms of KPM, the most time consuming process is the self-consistent calculation of Chebyshev coefficients
$\bm{\mu}_n^{\beta \alpha}$ of the leads. As a matter of fact, the requirement of including all details of leads into the calculation like this is a notoriously expensive cost in many quantum transport simulations\cite{WideBand2018,JHuangPhD,JTLu2014,Zelovich2015}. Now we will present two practical improvements of the algorithm.

\subsection{Chain Shaped Leads}

The first convenient simplification is to reduce the shape of leads into independent and semi-infinite 1D chains, as shown in Fig. \ref{FigImprovements} (a). Without transverse coupling in the lead, the coefficients are diagonal $\bm{\mu}_n^{\beta \alpha}=\delta_{\beta\alpha}\bm{\mu}_n^{\alpha \alpha}$ and they are identical for different $\alpha$. Now we only need to self-consistently calculate the Chebyshev coefficients of a 1D chain with width $W=1$, and the dimension of the matrix $\bm{R}$ in Eq. (\ref{EqR}) is reduced from $W \times J + W \times N$ to $M+N=J+N$. However, the mismatch between the leads and the conductor will give rise to additional scattering at their boundaries. Therefore, this brute simplification is mostly suitable for topological materials where backscattering have been prohibited by protections from topology and/or symmetry\cite{Bernevig2006}. See Examples B and C in the Section VI for simulation results.

\subsection{Finite Lead Approximation}

Here we propose another simple but efficient approximation to circumvent these difficulties.
The original setup was that both leads should be semi-\emph{infinitely} long, as illustrated in Fig. \ref{FigDevice}. Now we approximate both leads by two \emph{finite} ones, as presented in Fig. \ref{FigImprovements} (b). It is reasonable to imagine that the result will approach the correct one when their lengths $N_x^{L}$ and $N_x^{R}$ are sufficiently large. Now the conductor and leads are perfectly matched, so there will be no scattering on their boundaries.

If the lead lengths $N_x^{L}$ and $N_x^{R}$ needed to arrive within some precision are numerically acceptable,
this algorithm will be numerically more superior than the standard bath KPM described above.
For instance, the dressed Green's function can be obtained from Eq. (\ref{EqChebyshevGreen}) directly, only if $\bm{H}$ is the coupled Hamiltonian matrix of the whole system, the conductor \emph{and} two finite leads. Now, the sub-matrix $\bm{G}^{r(a)}_{ij}(E,H)$ (with $i,j$ running over the conductor sites) is naturally the approximation of the dressed Green's function (\ref{EqDressedGreenFunction})\cite{Datta}. In fact, due to the simple algebraic structure of Eq. (\ref{EqnTransmission}), one only needs the matrix indices $(i,j)$ running over the boundary sites connected to two leads. This process avoids the construction and calculation of complicated and non-Hermitian matrices like $R$ defined in Eq. (\ref{EqR}). A non-Hermitian matrix has complex eigenvalues, leading to difficulties of scaling itself with Eq. (\ref{equ:A9}) by its maximum and minimum eigenvalues, but an appropriate scaling of the matrix is key in the context of KPM.

Similarly, the surface Green's function of the lead can also be approximated by that of the finite one from Eq. (\ref{EqChebyshevGreen}), then the self energy is calculated with the help of Eq. (\ref{EqSelfEnergy}). In brief, this method circumvents all complicated steps of the standard bath KPM described in Section IV, especially the self-consistency calculation of the self energy, which is very time-consuming.

\section{Examples }

In this section, we present results from above KPM to calculate the two-terminal conductance of some example models.

\subsection{Square Lattice Conductor with Square Lattice Leads}

The first example is the two-dimensional square lattice with nearest hopping $t$,
\begin{equation}\label{EqHamiltonianSquareLattice}
H=\sum_{\langle\alpha,\beta\rangle} t c_{\alpha}^{\dagger}c_{\beta},
\end{equation}
where $\alpha$ and $\beta$ are indices for sites (each with only one spinorbital) in the conductor and leads, and $\langle\alpha,\beta\rangle$ run over all nearest site pairs.
The size of the conductor is $L\times W$, and the widths of both leads are also $W$.

The results from the standard bath KPM are shown in Fig. \ref{FigDifference}. Here the transmission $T$ a function of energy $E$ is plotted as the solid lines, for conductor sizes: (a) $25\times 25$, (b) $60\times 60$ (b) and (c) $100\times 100$. Different line colors corresponding to different Chebyshev terms $N$ are also shown in each panel. As a comparison, the red dashed line is the result from a direct matrix calculation of Eq. (\ref{EqnTransmission}), without using KPM. Without any disorder in the conductor, the conductance is quantized as plateaus with values $p\frac{e^2}{h}$, with $p$ the number of active channels at energy $E$\cite{Datta}. Smaller $N$ effectively corresponds to a stronger dephasing\cite{KPMRMP,LeiLiu2018}, and therefore the conductance is not perfectly quantized. The largest deviation at smaller $N$ (green lines in Fig. \ref{FigDifference}) happens around the band center $E=0$, which is a van Hove singularity.
This enhanced scattering is caused by the extremely large density of states around such a singularity\cite{Economou}.

On the other hand, larger conductor sizes need more Chebyshev terms to reach the perfect conductance value of quantum transport. This is understandable since a larger conductor gives rise to a longer journey for the electron to experience the dephasing, which is induced by the finiteness of Chebyshev terms $N$. Due to the tedious process of the standard bath KPM, especially the self-consistent calculation of the self energies, it is even more time-consuming than the direct matrix calculation.

\subsection{Square Lattice Conductor with Chain Shaped Leads}

\begin{figure}[htbp]
\centering
\includegraphics*[width=0.4\textwidth]{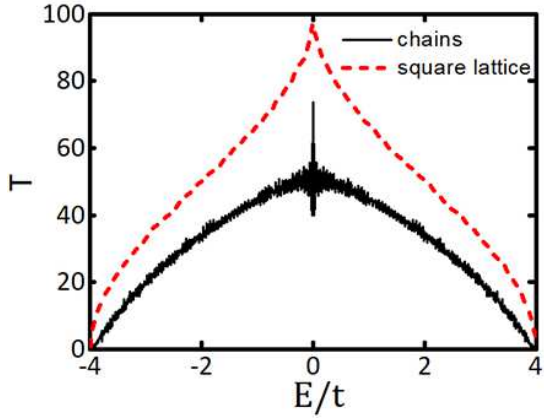}
\centering
\caption{ The black line is the result for Example B, square lattice conductor with chain shaped leads as illustrated in Fig. \ref{FigImprovements} (a). The conductor size is $100\times 100$, and the number of Chebyshev polynomial terms is $N=10000$. The red line is identical to that in Fig. \ref{FigDifference} (c), the result of square lattice leads (from direct matrix calculation), as a reference.  }
\label{FigChainLead}
\end{figure}

The results are shown as the black solid line of Fig. \ref{FigChainLead} (b), where the result of square lattice leads (red dashed line) are also presented as a comparison. Similar to the popular method of wide band approximation\cite{WideBand2018,JHuangPhD,Haug2008}, this simplification largely reduces the time and space consumptions of the self-consistent calculation of the lead self energy. On the other hand, the mismatch between the lead and the conductor will give rise to remarkable additional scattering on the interface, leading to a distinct reduction of the conductance compared with the case of perfect leads.

\subsection{Topological Material Conductor with Chain Shaped Leads}

\begin{figure}[htbp]
\centering
\includegraphics*[width=0.35\textwidth]{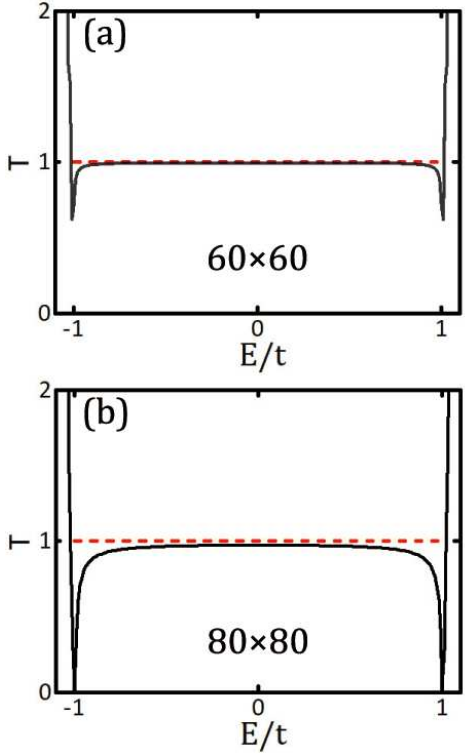}
\centering
\caption{ Black solid lines are results for Example C, topological material conductor with chain shaped leads. The conductor is the quantum anomalous Hall model defined in Eqs. (\ref{EqBHZ1}) and (\ref{EqBHZ2}), and the chain shaped leads are illustrated in Fig. \ref{FigImprovements} (a). The red dashed lines mark the quantized conductance value in units of $\frac{e^2}{h}$.
(a) Conductor size $60 \times 60$, Chebyshev terms $N=17000$.
(b) Conductor size $80 \times 80$, Chebyshev terms $N=9000$. The model parameters are: $A=1$, $B=-1$ ,$C=D=0$, and $M=-2$. The hopping in the leads, and the coupling hopping between the conductor and the lead is $t=1$. }
\label{FigBHZ}
\end{figure}

However, such scattering will be practically avoided if the conductor is a topological material with robust transport against backscattering, or three dimensional conductor with sufficiently large number of transport channels. Therefore this method is most applicable in these contexts. Here we adopt the typical model of the quantum anomalous Hall effect, the spin-up component of the Bernevig-Hughes-Zhang (BHZ) model\cite{Bernevig2006}, defined on a two-orbital square lattice. The Hamiltonian in the $k$ space can be written as\cite{Bernevig2006}
\begin{equation}\label{EqBHZ1}
H=\sum_{\bm{k}}h_{\alpha\beta}(\bm{k})c^{\dagger}_{\bm{k}\alpha},
c_{\bm{k}\beta}
\end{equation}
where $h_{\alpha\beta}(\bm{k})$ is a $2\times2$ matrix defined as
\begin{eqnarray}\label{EqBHZ2}
h(\bm{k}) &=&d_{0}I_{2\times 2}+d_{1}\sigma _{x}+d_{2}\sigma
_{y}+d_{3}\sigma _{z}  \label{EqBHZ2} \\
d_{0}(\bm{k}) &=&-2D\big(2-\cos k_{x}-\cos k_{y}\big)  \notag \\
d_{1}(\bm{k}) &=&A\sin k_{x},\quad d_{2}(\bm{k})=-A\sin k_{y}  \notag \\
d_{3}(\bm{k}) &=&M-2B\big(2-\cos k_{x}-\cos k_{y}\big),  \notag
\end{eqnarray}
with $\sigma_{x,y,z}$ the Pauli matrices acting on the space of two orbitals.
The Chern number of this model is 1 when $B/M>0$, so that there will be a pair of topological edge states in the bulk gap $(-\frac{M}{2},\frac{M}{2})$. Due to the topological origin, the edge states will contribute a quantized conductance $1\times \frac{e^2}{h}$ that is robust against elastic backscattering.

Fig. \ref{FigBHZ} is the numerical results (black lines) of this topological model from our Chebyshev approach, with chain shaped leads as illustrated in Fig. \ref{FigImprovements} (a).
Panels (a) and (b) are for conductor sizes $60 \times 60$ and $80 \times 80$, respectively, and the red lines mark the reference position of the quantized conductance. We can see that in most of bulk gap region, the simulated conductance is perfectly consistent with that predicted by the topological invariant theory. For example in Fig. \ref{FigBHZ} (a), the numerical match can be larger than $99.5\%$ near the gap center $E=0$. As in previous examples, larger conductor sizes needs more Chebyshev terms to reach the perfect quantum transport value. Moreover, the transport near gap edges are more sensitive to scattering, which is a natural consequence of bulk-edge mixing\cite{BulkEdge}.

\subsection{Square Lattice Conductor with Finite Lead Approximation}
\begin{figure*}[htbp]
\centering
\includegraphics*[width=1.0\textwidth]{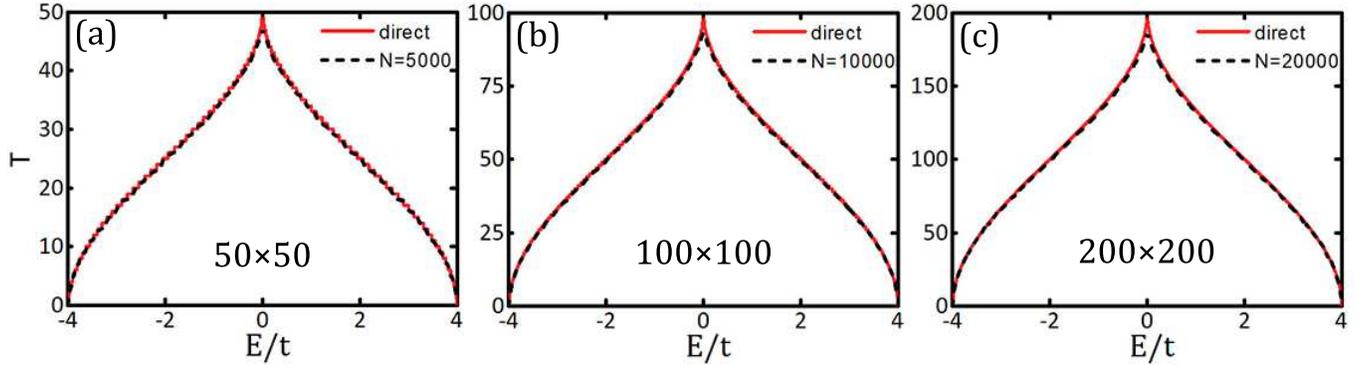}
\centering
\caption{Similar to Fig. \ref{FigDifference}, but the black solid lines are for Example D, with finite square lattice leads introduced in Section V B. The lengths of the leads are taken to be: (a) $40\times L$, (b) $30\times L$, and (c) $25\times L$. The number of Chebyshev terms are: (a) $N=5000$, (b) $N=10000$, and (c) $N=20000$. Red dashed lines are the results of a direct matrix evaluation of Eq. (\ref{EqnTransmission}) without KPM, as a reference. }
\label{FigFiniteDevice}
\end{figure*}

In Fig. \ref{FigFiniteDevice}, we present the results from the finite lead approximation, as introduced in Section V B.
Typically, the necessary length of the finite lead is less than 50 times of the conductor length, $N_x^{L},N_x^{R}\lesssim 50\times L$, to achieve a relative error less than $2\%$ throughout most of the energy spectrum. This necessitates a sparse matrix with dimension $\lesssim 100\times L\times W$ to store the Hamiltonian of the conductor \emph{and} leads. This matrix is structurally simpler, and usually not larger than $\bm{R}$ [Eq. (\ref{EqR2})] in the standard bath KPM, with an $N$ dependent dimension $(2N+L)\times W$ ($N$ is the number of Chebyshev terms, typically $\sim 10^3-10^4$) . Moreover, the calculation of self energies is also much simpler and faster than in the
standard bath KPM, since it is now a one-time process and no self-consistent loops are needed here.

\begin{figure}[htbp]
\centering
\includegraphics*[width=0.45\textwidth,bb=30 200 495 560]{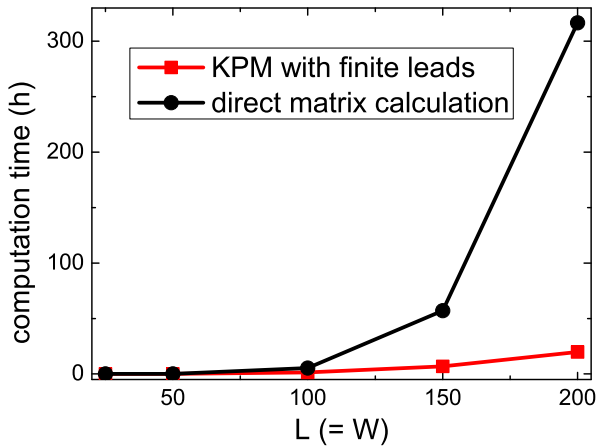}
\centering
\caption{ The computation times of obtaining a set of $T$ curve (by scanning over 800 energy points) as functions of the length of a square conductor $L$, by using the KPM with finite leads (red line and square dots), and by using the direct matrix evaluation of Eq. (\ref{EqnTransmission}) without KMP (black line and circle dots).  }
\label{FigTime}
\end{figure}

Moreover, here we will show that this method is also much faster than the direct matrix evaluation of Eq. (\ref{EqnTransmission}) without KMP. In Fig. \ref{FigTime}, the computation time of these two methods are plotted as functions of the length of the square shaped conductor. Here the ``computation time'' means the full time of obtaining one curve like those in Fig. \ref{FigFiniteDevice}, by calculating transmissions over 800 energy points. It can be seen that the KPM with finite leads is numerically much more advantageous than the traditional direct matrix calculation. As has been discussed above, since the Chebyshev coefficients have contained information on the full energy spectrum, this improved KPM method will be even more superior when one needs data on more energy points. Furthermore, in the context of simulating irregular shaped conductors with an irregular structure of Hamiltonian matrix, since the calculation cannot be reduced to a layer-to-layer recursive one\cite{QuantumTransport,Recursive}, this method we propose will be very applicable.

\section{Summary and Outlook}
In summary, we introduced the Chebyshev polynomial method to the Landauer-B\"{u}ttiker formula of the two-terminal transport device, by a generalization of the standard bath technique of KPM.
In this formula, the dressed Green's function can be expressed as Chebyshev polynomials of
the matrix $\bm{R}$ defined in Eq. (\ref{EqR}) or Eq. (\ref{EqR2}), and the self energies can be calculated through the dressed Green's function in a self-consistent way. During this process, the most resource consuming step is the calculation of self energies of the leads. A simple solution is to reduce the topology of the leads to parallel and decoupled atomic chains, but the price is additional scattering on the interfaces. Another solution is to approximate the leads as finite ones with sufficient length. This algorithm avoids complicated matrix calculations in the standard bath KPM (especially the self-consistent process of obtaining self energies), and also avoids notable boundary scattering in the chain shaped lead method. The numerical experiments verified that this method has a much less numerical cost than that of the traditional method of direct matrix calculation without KPM.

Since the leads themselves are not the object of study and there is a wide freedom of choosing leads. One of the future efforts is to find an appropriate design of leads or lead-conductor coupling\cite{JHuangPhD,JTLu2014,Zelovich2015,LeadGeometry}, with small resource demanding in the Chebyshev polynomial representation, while with small backscattering on the interfaces to the conductor. Furthermore, our method can also be generalized to Cheyshev forms to linear responses of other degree of freedoms of quantum transports structures\cite{Thermal2,Optical,PhononTransport,Mosso2019}.

\begin{acknowledgments}
We thank Prof. S.-J. Yuan (Wuhan University) for beneficial discussions. This work was supported by National Natural Science
Foundation of China under Grant Nos. 11774336 and 61427901. YYZ was also supported by the Starting Research Fund from Guangzhou University under Grant No. RQ2020082.
\end{acknowledgments}

\section{DATA AVAILABILITY}
The data that support the findings of this study are available from the corresponding author upon reasonable request.

\nocite{*}
\bibliography{aipsamp}

\end{document}